\documentclass[preprint]{aastex}

\begin{document}

\title{Evidence for an Additional Heat Source in the Warm Ionized Medium
of Galaxies}

\author{R. J. Reynolds} \author{L. M. Haffner} \and \author{S. L. Tufte}
\affil{Department of Astronomy, University of Wisconsin--Madison}
\affil{475 North Charter Street, Madison, WI 53706}
\email{reynolds@astro.wisc.edu, haffner@astro.wisc.edu, 
tufte@astro.wisc.edu}

\begin{abstract}

Spatial variations of the [S~II]/H$\alpha$ and [N~II]/H$\alpha$ line
intensity ratios observed in the gaseous halo of the Milky Way and other
galaxies are inconsistent with pure photoionization models.  They appear
to require a supplemental heating mechanism that increases the electron
temperature at low densities n$_e$.  This would imply that in addition to
photoionization, which has a heating rate per unit volume proportional to
n$_{e}^{2}$, there is another source of heat with a rate per unit volume
proportional to a lower power of n$_e$.  One possible mechanism is the
dissipation of interstellar plasma turbulence, which according to Minter
\& Spangler (1997) heats the ionized interstellar medium in the Milky Way
at a rate $\sim 1 \times 10^{-25}$n$_e$ ergs cm$^{-3}$ s$^{-1}$. If such
a source were present, it would dominate over photoionization heating in
regions where n$_e$ $\lesssim$ 0.1 cm$^{-3}$, producing the observed
increases in the [S~II]/H$\alpha$ and [N~II]/H$\alpha$ intensity ratios at
large distances from the galactic midplane, as well as accounting for the
constancy of [S~II]/[N~II], which is not explained by pure
photoionization.  Other supplemental heating sources, such as magnetic
reconnection, cosmic rays, or photoelectric emission from small grains,
could also account for these observations, provided they supply to the
warm ionized medium $\sim$ 10$^{-5}$ ergs s$^{-1}$ per cm$^2$ of Galactic
disk.

\end{abstract}


\section{Introduction}
\label{sec:intro}

Although the warm ionized medium (WIM), also called the diffuse ionized
gas (DIG), is a principal component of the interstellar medium in our
Galaxy and others, the source of its ionization and heating is not
understood (e.g., Reynolds 1995; Rand 1998).  Observed line intensities,
particularly the high values of [S~II]/H$\alpha$ and [N~II]/H$\alpha$
compared to those in traditional, discrete H II regions surrounding O and
early B stars suggest that photoionization by a dilute radiation field
plays an important role (e.g., Domg\"orgen \& Mathis 1994); both models
and observations indicate that these emission lines originate primarily
from warm ($\sim 10^4$ K) regions in which the hydrogen is nearly fully
ionized (e.g., Sembach et al 1999; Reynolds et al 1998).  It has been
suggested by Miller \& Cox (1993) and Dove \& Shull (1994), for example,
that Lyman continuum radiation originating from O stars penetrates the H I
cloud layer and ionizes diffuse interstellar gas within the disk and lower
halo.  While O stars are the only known source with sufficient power to
maintain the WIM, the high opacity of the interstellar H I has led others
to propose the existence of more widely distributed sources of ionization
(e.g., Slavin et al 1993; Mellott et al 1988 and Sciama 1990; Raymond
1992; Skibo \& Ramaty 1993).

\section{Problems with Pure Photoionization Models}
\label{sec:pro}

Photoionization models incorporating a low ionization parameter U (the
ratio of photon density to gas density) have been generally successful in
accounting for the elevated [S~II]/H$\alpha$ and [N~II]/H$\alpha$ and low
[O~III] $\lambda$5007/H$\alpha$ ratios observed in the WIM (e.g.,
Domg\"orgen \& Mathis 1994; Greenawalt, Walterbos, \& Braun 1997; Martin
1997; Wang, Heckman, \& Lehnert 1998). However, the models have failed to
explain observed \emph{variations} in some of the ratios.  For example,
Rand (1998) observed that [S~II]/H$\alpha$ and [N~II]/H$\alpha$ increase
with increasing distance $|$z$|$ from the midplane of NGC 891, having
values of 0.2 and 0.35, respectively, near z = 0, and 0.6 and 1.0,
respectively, near $|$z$|$ = 2000 pc.  To account for such large ratios at
high $|$z$|$, Rand had to adopt a hard stellar spectrum (an upper IMF
cutoff of 120 M$_{\odot}$) plus additional hardening as the radiation
propagated away from the midplane.  However, a hard spectrum appears to be
inconsistent with He~I $\lambda$5876 recombination line observations (Rand
1998, 1997, and references therein).

More significantly, the models fail to account for the fact that, while
the variations in [S~II]/H$\alpha$ and [N~II]/H$\alpha$ are large,
[S~II]/[N~II] remains nearly constant.  A similar behavior for [S~II],
[N~II], and H$\alpha$ has been observed in other galaxies (e.g., Golla,
Dettmar, \& Domg\"orgen 1996; Otte \& Dettmar 1999) as well as in the
Milky Way (Haffner, Reynolds, \& Tufte 1999).  Golla et al (1996) and Rand
(1998) have pointed out that the constant value of [S~II]/[N~II] cannot be
reproduced by photoionization models, because in these models variations
in [S~II]/H$\alpha$ and [N~II]/H$\alpha$ are primarily the result of
variations in the ionization parameter U, which always produce larger
changes in [S~II]/H$\alpha$ than in [N~II]/H$\alpha$. This is due to the
different ionization potentials of S and N, with the result that sulfur
can be primarily S$^+$ or primarily S$^{++}$, depending on the spectrum
and strength of the radiation field, whereas nitrogen remains
primarily N$^+$ under nearly all WIM conditions (e.g., Howk \& Savage
1999).

Another observation that pure photoionization models fail to reproduce is
the rise in [O~III]/H$\beta$ with increasing $|$z$|$, or increasing
[S~II]/H$\alpha$ and [N~II]/H$\alpha$ (Rand 1998, Greenawalt et al 1997).
In NGC 891, for example, the [O~III]/H$\beta$ intensity ratio more than
doubles from 0.3 at z = 0 to about 0.75 a $|$z$|$ = 2000 pc (Rand 1998).  
The models predict the opposite trend.  Rand proposed an additional source
of collisional ionization at high $|$z$|$ to account for the enhanced
[O~III] intensity, but emphasized that this would still not explain the
constancy of the [S~II]/[N~II] ratio.

We show that these line ratio variations could be explained by the
existence of an additional source of heat in the diffuse ionized gas.  In
the following sections we use recent emission line data for the Milky Way,
obtained with the Wisconsin H-Alpha Mapper (WHAM) spectrometer, to derive
the required heating rates and place constraints on possible supplemental
heating mechanisms.

\section{Line Ratio Variations Due to an Additional Heat Source}
\label{sec:var}

\subsection{Evidence for Variations in Electron Temperature}
\label{sec:evi}

Haffner et al (1999) have shown that these emission line observations can
be readily explained if the large variations in [N~II]/H$\alpha$ and
[S~II]/H$\alpha$ are due primarily to variations in the electron
temperature T$_e$ rather than to variations in the ionization parameter U.  
The constancy of [S~II]/[N~II] is then a consequence of the fact that the
two lines have nearly the same excitation energy.  Such temperature
variations could also produce increases in [O~III]/H$\beta$, perhaps
eliminating the need for a secondary source of ionization.  For the Milky
Way, Haffner et al (1999) found that an increase in T$_e$ from 7000 K at
$|$z$|$ = 500 pc to approximately 10,000 K at 1500 pc would produce the
observed factor of three increases in the [N~II]/H$\alpha$ and
[S~II]/H$\alpha$ ratios while keeping [S~II]/[N~II] constant.  Elevated
temperatures have also been proposed by Bland-Hawthorn, Freeman, \& Quinn
(1997) to account for the anomalously high [N~II]/H$\alpha$ in the diffuse
gas at the outer edge of the disk of NGC 253.  They concluded that the
high ratio could not be explained by photoionization alone, but required
an additional heat source at large galactic radius that would
``selectively heat the electrons without producing a higher ionization
state of nitrogen.''

These emission line data suggest that the regions with higher
[S~II]/H$\alpha$ and [N~II]/H$\alpha$ ratios (i.e., higher temperatures)
are regions not just at larger distances $|$z$|$ from the galactic
midplane, but more generally are regions with lower electron density.  
This is indicated by the strong anticorrelation between these line ratios
and the H$\alpha$ intensity.  This anticorrelation is apparent not only in
the data showing increasing ratios with increasing $|$z$|$, but also in
observations at constant $|$z$|$ (e.g., Rand 1998; Otte \& Dettmar 1999;
Domg\"orgen \& Dettmar 1997; Golla et al 1996; Ferguson, Wyse, \&
Gallagher 1996) and at large galactocentric radii (Bland-Hawthorn et al
1997).  A strong anticorrelation is also found in the observations of the
Milky Way (Haffner 1999; Haffner et al 1999), again, not only with
increasing $|$z$|$, but also for lines of sight that sample just the
relatively low $|$z$|$ gas in the local Orion arm.  Since it is difficult
to see how the integration length could affect the temperature, we
conclude that variations that are correlated with H$\alpha$ intensity
(i.e., emission measure) are actually variations correlated with density.

If the temperature in fact varies inversely with density in the diffuse
ionized gas, then there must be an additional heat source that dominates
over ionization heating at low densities.  The heating rate per unit
volume from photoionization is limited by recombination and is thus
proportional to n$_{e}^{2}$.  The cooling rate per unit volume depends
upon electron-ion collisions and is also proportional to n$_{e}^{2}$.  
Therefore, with only photoionization, T$_e$ is nearly independent of n$_e$
(although the density dependence of the ion ratios will have some effect
on the equilibrium temperature).  However, if an additional heating term
were added that was proportional to n$_e$, or did not depend upon density
at all, it would dominate at sufficiently low densities, increasing the
equilibrium temperature and producing an inverse relationship between
T$_e$ and n$_e$ (Reynolds \& Cox 1992).  This additional heating term
would decouple the heating of the gas from its ionization, driving up the
the intensities of [S~II] and [N~II] relative to H$\alpha$ while not
affecting the ionization states of S and N, i.e., allowing the
[S~II]/[N~II] ratio to remain constant.  Such heat sources in the WIM may
include, for example, photoelectric heating by dust, the dissipation of
interstellar turbulence, and coulomb collisions with cosmic rays, which
are proportional to n$_e$ (Draine 1978; Minter \& Balser 1997; Skibo,
Ramaty, \& Purcell 1996), and magnetic field reconnection, which may be
nearly independent of density (Gon\c{c}alves, Jatenco-Pereira, \& Opher
1993).

\subsection{Electron Temperature vs $|$z$|$ in the Perseus Arm}
\label{sec:ele}

The heating rates due to both photoionization and the supplemental source
can be estimated by fitting the predicted variation in [N~II]/H$\alpha$
with temperature to the observed variation in this line ratio.  This can
be done for the Perseus spiral arm of the Milky Way, where the associated
optical emission lines have been kinematically identified and observed to
high Galactic latitude with the WHAM spectrometer (Haffner 1999; Haffner
et al 1999). As a result, these Perseus arm observations provide both line
intensity ratios and electron densities as a function of distance from the
Galactic midplane.  Figure 1 presents the electron temperatures T$_e$ vs
$|$z$|$ derived from these [N~II]/H$\alpha$ data and the relationship
between T$_e$ and [N~II]/H$\alpha$ given by

\begin{equation}
\frac{I_{[\mathrm{N II}]}}{I_{\mathrm{H}\alpha}} = 1.63 \times 10^5 \,
\left( \frac{\mathrm{N}^+}{\mathrm{N}} \right) \, \left(
\frac{\mathrm{H}^+}{\mathrm{H}} \right)^{-1} \, 
\left( \frac{\mathrm{N}}{\mathrm{H}} \right) \, T^{0.426}_{4} \,
e^{-2.18/T_4}, 
\end{equation} \\
where $T_4$ is T$_e$ in units of 10$^4$ K, N/H is the gas phase abundance
of nitrogen, and N$^+$/H and H$^+$/H are the fraction of nitrogen and
hydrogen, respectively, that is singly ionized. Since N$^+$/N $\approx$
H$^+$/H (e.g., Howk \& Savage 1999; Haffner et al 1999) and N/H $\simeq$
$7.5 \times 10^{-5}$ (Meyer, Cardelli, \& Sofia 1997), equation (1) can be
rewritten simply as

\begin{equation}
\frac{I_{[\mathrm{N II}]}}{I_{\mathrm{H}\alpha}} = 12.2 \, T^{0.426}_{4}
\, e^{-2.18/T_4}.
\end{equation}

Equation (2) and the plot of [N~II]/H$\alpha$ vs Galactic latitude
presented in Figure~8 of Haffner et al (1999) were then combined to
produce the T$_e$ vs $|$z$|$ relation for the Perseus arm in Figure 1.
This result covers the Galactic latitude range $-34^{\rm o} \geq b \geq
-6^{\rm o}$ averaged over the longitude interval $125^{\rm o} \geq \ell
\geq 152^{\rm o}$.  The distance to the Perseus arm is assumed to be 2.5
kpc (Reynolds et al 1995, and references therein).

\subsection{Electron Density vs $|$z$|$ in the Perseus Arm}
\label{sec:den}

     The electron density n$_e$ within the WIM at a distance $|$z$|$ from
the midplane can be derived from the H$\alpha$ intensity, which is related
to the emission measure EM through the relation EM = 2.75 $T^{0.9}_{4}
I_{\mathrm{H}\alpha}$ (from Martin 1988), where EM (= $\int
\mathrm{n}_{e}^{2}$ ds) is in units of cm$^{-6}$ pc and $I_{H\alpha}$ is
in rayleighs (1 R = 10$^6$/4$\pi$ photons cm$^{-2}$ s$^{-1}$ sr$^{-1}$).  
Along the line of sight through the Perseus arm, EM can be expressed as
n$^{2}_{e}f$L, where L is the path length through the arm, $f$ the
fraction of L occupied by ionized hydrogen, and n$_e$ the rms electron
density within the ionized regions. Haffner et al (1999) showed that for
the same ranges of $\ell$ and $b$ represented in Figure 1,
$I_{\mathrm{H}\alpha}$($|$z$|$) $\simeq$ 5.7 e$^{-|\mathrm{z}|/500}$ R.
Therefore, if L is assumed to have a value of 1000 pc (see Fig. 1 in
Becker \& Fenkart 1970),

\begin{equation}
\mathrm{n}_e(|\mathrm{z}|) = 0.125 \, T^{0.45}_{4} f^{-0.5} \,
e^{-|\mathrm{z}|/1000} \; \mathrm{cm}^{-3}.
\end{equation}

We consider two situations: 1) a constant filling fraction $f$ = 0.2
(Reynolds 1991), and 2) a filling fraction that increases with $|$z$|$
according to the relation given by Kulkarni \& Heiles (1987), that is,
$f(|\mathrm{z}|) = 0.1 \, e^{|\mathrm{z}|/750}$ for $|\mathrm{z}| < 1740$
pc.  There is some evidence that $f$ does in fact increase with distance
from the midplane (Reynolds 1991); however, the results are sufficiently
uncertain that both a constant and varying $f$ are considered here.
     
\subsection{Fits to the T$_e$ vs $|$z$|$ Relation}
\label{sec:fit}

We assume that the temperature is determined by a balance between the
cooling rate per unit volume ($\Lambda$n$^{2}_{e}$) in the diffuse
ionized gas and two heating rates: the net heating by photoionization,
given by G$_{0}$n$^{2}_{e}$, plus an additional heating term, given
either by G$_1$n$_e$ or by just a constant G$_2$.  The heating--cooling
balance can then be expressed as either $\mathrm{G}_{0} +
\mathrm{G}_{1}/\mathrm{n}_{e} = \Lambda$, or $\mathrm{G}_{0} +
\mathrm{G}_{2}/\mathrm{n}^{2}_{e} = \Lambda$, representing, for example,
supplemental heating by turbulent dissipation (G$_{1}$), or by magnetic
field reconnection (G$_{2}$), respectively.  Therefore, depending upon the
values of G$_{1}$ or G$_{2}$ relative to G$_{0}$, significantly increased
heating (relative to photoionization) can occur as the density decreases.  
This will result in higher equilibrium temperatures at lower densities.

We have adopted the cooling function $\Lambda$ for low density
photoionized gas given in Osterbrock (1989).  While this particular
cooling function may not be exactly appropriate for the WIM, for T$_e \,>$
7000 K (the temperature range considered here) it is a very good
approximation, because like the model H II region in Osterbrock, the WIM's
cooling function is dominated by [O~II] and [N~II].  An equilibrium
temperature for each value of $|$z$|$ can be computed from equation (3)
and one of the above heating--cooling balance equations.  The values of
G$_{0}$ and G$_{1}$, or G$_{0}$ and G$_{2}$, can then be adjusted to fit
the T$_e$ vs $|$z$|$ distribution in Figure 1.  The best-fit values are
listed in Table 1 for four cases: (a) supplemental heating G$_{1}$n$_e$
and constant $f$; (b) supplemental heating G$_{2}$ and constant $f$; (c)
supplemental heating G$_{1}$n$_e$ and a variable $f(|$z$|$); and (d)
supplemental heating G$_{2}$ and variable $f(|$z$|$).  The associated
best-fit curves are also plotted on Figure 1 for comparison with T$_e$ vs
$|$z$|$ inferred from the observed [N~II]/H$\alpha$ ratios in the Perseus
arm.  Note that the derived values for G$_{1}$ and G$_{2}$ are
proportional to L$^{-\case{1}{2}}$ and L$^{-1}$, respectively, where L is
the assumed path length through the Perseus arm.  Also, Haffner et al
(1999) discussed the possible contamination of the [N~II] spectra by a
weak atmospheric emission line. If this line is present with the intensity
of their upper limit (0.1 R), then the best fit values for G$_{1}$ and
G$_{2}$ would be 20\% -- 30\% lower than those presented in Table 1, while
G$_{0}$ would be less affected.

\section{Discussion}
\label{sec:dis}

Figure 1 shows that all four cases give good fits to the inferred T$_e$ vs
$|$z$|$ distribution, within the uncertainty implied by the jaggedness of
the distribution.  Therefore, a supplemental heat source with a heating
rate per unit volume proportional to n$_{e}^{1}$ or n$_{e}^{0}$ could
account for the observed variations in the line intensity ratios.  
Moreover, these results place tight constraints on the required rates,
implying a photoionization heating rate coefficient G$_0$ $\approx 1
\times 10^{-24}$ ergs cm$^{+3}$ s$^{-1}$, and a supplemental rate
coefficient of either G$_1$ $\sim 1 \times 10^{-25}$ ergs s$^{-1}$ or
G$_2$ $\sim$ few $\times 10^{-27}$ ergs cm$^{-3}$ s$^{-1}$.  Thus for
n$_e$ greater than 1 cm$^{-3}$, the heating rate per unit volume is
dominated by photoionization, while below 0.1 -- 0.04 cm$^{-3}$, the
supplemental heating dominates.  This value for G$_0$ corresponds to a
stellar ionizing spectrum with T$_{eff}$ $\approx$ 30,000 -- 35,000 K
(Osterbrock 1989), i.e., late O to early B, and is consistent with the
observations of weak He I recombination line emission from the WIM (Tufte
1997, Reynolds \& Tufte 1995; Heiles et al 1996).

Values of G$_1$ near 1 $\times 10^{-25}$ ergs s$^{-1}$ (Table 1) have in
fact been predicted for the WIM in the Milky Way by models of
photoelectric grain heating (Reynolds \& Cox 1992; Draine 1978) and by
models of the dissipation of interstellar turbulence (Minter \& Spangler
1997).  At electron temperatures above 8000 K the net heating by grains
decreases sharply due to cooling by electron--grain collisions (Draine
1978), and, therefore, this process is not likely to account for the
9,000~K -- 11,000~K temperatures at high $|$z$|$, unless photoelectric
heating in the WIM is dominated by large molecules (e.g., PAHs) (Lepp \&
Dalgarno 1988). Minter \& Spangler (1997), on the other hand, have
predicted an energy dissipation rate of approximately 1 $ \times
10^{-25}$n$_e$ ergs cm$^{-3}$ s$^{-1}$ due to ion--neutral collisional
dampening in the Milky Way's nearly fully ionized 10$^4$ K WIM. They
concluded that the dissipation of turbulence probably plays a major role
in heating the WIM and contributing to the [S~II] and [N~II] emission (see
also Minter \& Balser 1997 and Tufte, Reynolds, \& Haffner 1999).  
Another potential source is coulomb collisions by cosmic rays, which
according to some interpretations of the Galactic $\gamma$-ray background,
could deposit significant power into the interstellar gas (e.g., Skibo et
al 1996; Valinia \& Marshall 1998).
 
A heating mechanism that is independent of density (curves b and d in Fig.
1) could also account for these temperature variations.  One such
mechanism is magnetic field reconnection (Raymond 1992; Birk, Lesch, \&
Neukirch 1998; Gon\c{c}alves et al 1993).  A field strength as high as 7
$\mu$G (Webber 1998; Heiles 1995) and a time scale of 10$^8$ yr for the
amplification of the field by the Galactic dynamo (Raymond 1992, and
references therein) would provide an average power of 6 $\times 10^{-28}$
erg cm$^{-3}$ s$^{-1}$, a rate that is a factor of 2.5 to 13 below the
values for G$_2$ given in Table 1 ---but approximately the rate that is
needed, if reconnection occurred only within the more limited volume of
the WIM.

\section{Summary and Concluding Remarks}
\label{con}

The anticorrelation between H$\alpha$ intensity and the line intensity
ratios [N~II]/H$\alpha$ and [S~II]/H$\alpha$, as well as the constancy of
[N~II]/[S~II] in the diffuse ionized gas of the Milky Way and other
galaxies can be explained if the electron temperature T$_e$ increases with
decreasing density n$_e$.  An inverse relationship between T$_e$ and n$_e$
would imply that, in addition to photoionization, which heats at a rate
proportional to n$_{e}^{2}$, there is an additional source that is
proportional to a lower power of n$_e$.

In the Milky Way the dissipation of interstellar turbulence, with a
predicted rate $\sim 1 \times 10^{-25}$ n$_e$ ergs cm$^{-3}$ s$^{-1}$ in
the WIM (Minter \& Spangler 1998), may be the source of this additional
heating, raising the possibility that the observed increases in forbidden
line intensities (relative to H$\alpha$) in galactic halos is the final
step in a turbulent energy cascade that begins with large scale motions of
the interstellar gas.  However, other mechanisms, such as heating by PAHs,
magnetic reconnection, or cosmic rays are also possible provided that
within the WIM they have a rate coefficient of the magnitude listed in
Table 1.

Measurements of higher T$_e$ in regions with higher [N II]/H$\alpha$ and
[S~II]/H$\alpha$ would provide strong, independent support for the
existence of such supplemental heating.  These measurements could perhaps
be made through accurate observations of the H$\alpha$, [N~II], and [S~II]
line widths (e.g., Reynolds 1985), or through observations of other
emission lines such as [O~II] $\lambda$3727 and the extremely faint [N~II]
$\lambda$5755 line, which have higher excitation energies, and thus are
more temperature sensitive, than [N~II] $\lambda$6584 and [S~II]
$\lambda$6716 (see Ferguson et al 1996).  We hope to begin some of these
investigations in the near future.

This work was supported by the National Science Foundation grant AST96
19424.

\newpage

\begin{deluxetable}{cccc}
  \tablewidth{0pt}
  \tablecaption{Heating Rate Coefficients for the Perseus Arm}
  \tablehead{
    \colhead{Case} &
    \colhead{G$_0$} &
    \colhead{G$_1$} &
    \colhead{G$_2$} \\
    \colhead{} &
    \colhead{($10^{-24}$ ergs cm$^{+3}$ s$^{-1}$)} &
    \colhead{($10^{-25}$ ergs s$^{-1}$)} &
    \colhead{($10^{-27}$ ergs cm$^{-3}$ s$^{-1}$)}
    }

  \startdata

  a & $0.3$ & $1.6$ & \nodata \\
  b & $0.95$ & \nodata & $7.6$ \\
  c & $0.9$ & $0.65$ & \nodata \\
  d & $1.35$ & \nodata & $1.5$  
  
  \enddata
\end{deluxetable}

\newpage

\figcaption[fig1.ps]{Electron temperatures T$_e$ inferred from the
  [N~II]/H$\alpha$ line intensity ratios plotted vs the distance $|$z$|$
  from the Galactic midplane in the Perseus spiral arm (bold line).  Also
  plotted are the best fits to this T$_e$ vs $|$z$|$ relation for four
  cases in which the gas is heated by photoionization plus an additional
  non-ionizing source (see text).
  \label{fig:temp}}

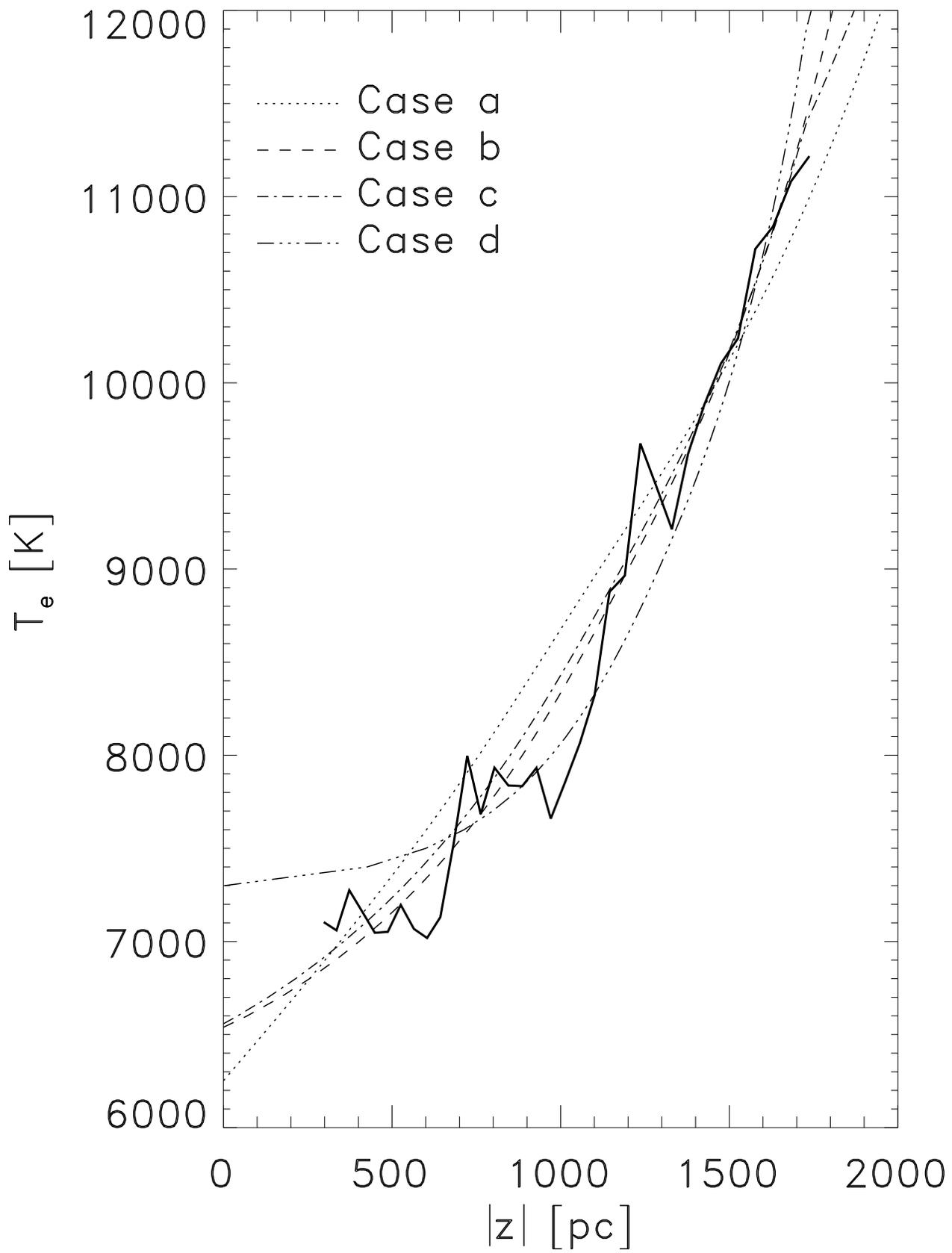
\end{document}